\documentclass[times,final]{elsarticle}
\usepackage{graphicx}
\usepackage{amssymb}
\usepackage{amsmath}

\journal{Physica A}

\biboptions{sort&compress}

\begin{document}
.
\begin{frontmatter}
\title{Equilibrium distributions in entropy driven balanced processes}
\author[label1]{Tam\'as~S.~ Bir\'o}
\author[label2]{Zolt\'an N\'eda}
\address[label1]{HIRG, HAS Wigner Research Centre for Physics, Budapest, Hungary} 
\address[label2]{Babe\c{s}-Bolyai University, Department of Physics 
1 Kog\u{a}lniceanu str., 400084 Cluj, Romania} 
\date{}

\begin{abstract}
For entropy driven balanced processes
we obtain final states with Poisson, Bernoulli, negative binomial and P\'olya distributions.
We apply this both for complex networks and particle production. 
For random networks we follow the evolution of the degree distribution,
$P_n$, in a system where a node can activate $k$ fixed connections from $K$ possible partnerships 
among all nodes.  The total number of connections, $N$, is also fixed. 
For particle physics problems $P_n$ is the probability of having
$n$ particles (or other quanta) distributed among $k$ states
(phase space cells) while altogether a fixed number of $N$
particles reside on $K$ states.
\end{abstract}

\end{frontmatter}
\newcommand{\vs}{\vspace{3mm}}

\newcommand{\be}{\begin{equation}}
\newcommand{\ee}[1]{\label{#1} \end{equation}}
\newcommand{\ba}{\begin{eqnarray}}
\newcommand{\ea}[1]{\label{#1} \end{eqnarray}}
\newcommand{\nl}{\nonumber \\}
\newcommand{\ave}{\overline{u}}
\newcommand{\exv}[1]{ \left\langle {#1} \right\rangle}

\newcommand{\sumi}{ \sum_{n=0}^{\infty}\limits\!}

\section{Introduction}

The general laws of equilibrium and near-equilibrium thermodynamics are 
classical knowledge. The existence of macro-equilibrium based
on micro-dynamics and its stability properties are closely connected with the
physical notions of temperature, heat and entropy.  
There are, however, still some open problems left to modern statistical physics.
Generalizations of the entropy -- probability 
connection\cite{RENYI1,NAZAROV1,TSALLIS-CLASSICAL,TSALLISBOOK,SUPERSTAT,QDIST,IDEALGAS}, 
beyond mathematical games, 
also require the re-interpretation of the notion of equilibrium\cite{ZEROTHLAW} 
and the composition rules for uniting smaller systems in bigger and more complex
ones\cite{ABSTRACTRULES,RULES}. 
Also questions, related to far from equilibrium behavior of large dynamical
systems, like growing 
networks\cite{ERDOS-RENYI,ERDOSRENYI1,WATTS-STROGATZ,BARABASI-SCIENCE,BARABASI-REVMOD,THURNER,HORVAT}, 
are intriguing.

Here we consider a unified approach to all statistics  resulting
from a balanced  micro-dynamics applicable to a wide class of physical models.  
In particular we discuss the case of  randomly connected networks and 
randomly produced particles with some imposed conservation laws. 

For processes near 
equilibrium typically a subsystem and a reservoir exchange physical currents
in a locally symmetric and microscopically reversible way, establishing in due of
time a detailed balance. This state is characterized then by the distribution
of those conserved quantities.
Our first example is the hadronization process:
In high-energy accelerator experiments the number of created particles per event fluctuates.
Since the total energy is fixed in such experiments, the distribution of
hadron numbers from one collision event to another determines 
the effective thermal-like  properties of the observed kinetic spectra, 
\cite{IDEALGAS,BIRO-SIGMAPHI2014}. 

Another example is given by random networks, where due to a balance between growth and 
decays of the connections the degree distribution tends to a few particular shapes in equilibrium.
Such studies have become popular in the last decades
\cite{KULLMAN-KERTESZ,KRAPIVSKY,KRAPIVSKY-MASTER,DOROG1,DOROG2,DOROG3}.
Random networks are characterized by the probability distribution, $P_n$,
of having a given number of links, $n$, known as the degree distribution. 
The connection between the indexed nodes, $i=0, 1,\ldots,k$, can be described by an
adjacency matrix, $C_{ij}$ containing zero for no connection and a number
for a link pointing from node $i$ to node $j$.
In unweighted networks the entries of $C_{ij}$ are just zeros and ones,
and for undirected networks only the upper triangle of the matrix is used. For more general
considerations, however, e.g. on directed networks this matrix is not
necessarily symmetric, $C_{ji}\ne C_{ij}$. Weighted connections
also may be of relevance for some statistical problems, in such cases
$C_{ij}$ can be any real number. Even self-connections,
$C_{ii}\ne 0$ have to be allowed for the most general network.

In this view a random network is a random ensemble of $C_{ij}$ values.
It looks analogous to a rectangular box with altogether $K = k \times k$ cells, onto which
$N$ (multiple, including self-) connections are randomly thrown.
By an analysis, in particular by asking for the probability of
a given multiplicity connection from a single node, one chooses to sum over
$k$ cells in a row (or in a column) and asks for the $P_n$ probability
for finding exactly $n$ connections. 
This is similar thus with the case of particle physics problems, where  
$P_n$ is the probability of having
$n$ particles (or other quanta) distributed among $k$ states
(phase space cells) while altogether a fixed number of $N$
particles reside on $K$ states.

In a general picture applicable both for particle and link distributions under some conservation constraints,
 we assume $K$ cells, $N=pK$ ''stones''
(i.e. units of connection strength). 
If only $0$ or $1$ stone can be in a cell then $p<1$ ({\em fermionic} systems), for an arbitrary 
number of multiple connections $p>1$ is also possible ({\em bosonic} systems). 
The stationary "degree distribution" of the nodes is given by
the probability that a single row (with $k$ boxes) contains {\em exactly} $n$ stones
$Q_n={\mathrm{Prob}}(n,k; N=pK, K=k^2)$.

We present analytic solutions to the above problems in the frameworks of i) a pure statistical
counting and ii) in a master equation approach.

\section{Statistics of random displacements}

For a totally random displacement of $N$ particles in $K$ cells and asking for observing
$n$ ones in $k$ cells one obtains the distribution following the idea suggested by Boltzmann:
 the probability of such an observation is given by the ratio of the numbers of arrangements with and 
 without splitting the system to $k$ and $K-k$ cells, respectively. 
 The number of  combinations of $n$ particles in $k$ cells, if each cell can be occupied at most by one 
 particle (the {\em fermionic} case), is given by:
\be
 W(k,n) \: = \: \frac{k!}{n! \, (k-n)!} \: = \: \binom{k}{n}.
\ee{COMBINUM}
In particle physics fermions behave this way and for networks this result corresponds to
unweighted directed links.
Networks constrained by $N \le K$ and $n \le k$ are {\em fermionic}.
If on the other hand $N > K$ and correspondingly $n > k$ is allowed, then there must be 
multiple connections. Such systems we label as {\em bosonic} ones.

The probability of having exactly $n$ particles (connections) in $k$ cells
(from maximal $k$ connections) while in a huge system altogether $N$ particles
(connections) are randomly distributed in $K$ cells (among the maximal number of partner nodes)
allowing only single occupation (single connections) 
is given by the following P\'olya distribution:
\be
 Q_n \: = \: \frac{W(k,n) \, W(K-k, N-n)}{W(K,N)} \: = \: 
 \frac{\binom{k}{n} \, \binom{K-k}{N-n}}{\binom{K}{N}}.
\ee{POLYA}
This distribution is normalized, $\sumi Q_n \, = 1$.
For accessing different limits of the P\'olya distribution, we utilize the generic
approximation
\be
 W(K,N) \quad \xrightarrow[{K \gg N}]{} \quad \frac{K^N}{N!},
\ee{COMBILIMIT}
and correspondingly get a Bernoulli distribution,
\be
Q_n \quad \xrightarrow[K \gg N]{k \gg n} \quad \binom{N}{n} \,
\left(\frac{k}{K} \right)^n \, \left(1-\frac{k}{K} \right)^{N-n}.
\ee{BERN1}
On the other hand, considering the small subsystem limit, $n \ll N$ and $k \ll K$,
in eq.(\ref{POLYA}) we arrive at a complementary Bernoulli distribution,
\be
Q_n \quad \xrightarrow[K \gg k]{N \gg n} \quad \,
 \binom{k}{n} \, \left(\frac{N}{K} \right)^n \, \left(1-\frac{N}{K} \right)^{k-n}.
\ee{BERN2}
We note that for both distributions $\exv{n}=Nk/K=kN/K$, but the variance and other higher
moments differ.
They both lead to a Poisson distribution with the above  parameter $\exv{n}$ in the
$k\to\infty$ limit:
\be
Q_n \quad \xrightarrow[N \gg n, \:  kN/K=\exv{n}]{K \gg k \gg n} \quad 
\frac{\exv{n}}{n!}  \, e^{-\exv{n}}.
\ee{POISSON}


We discuss briefly the classical case of distinguishable particles, where the
particles have identities. In such a case the total number of arrangements of 
the $N$ particles in $K$ cells is simply $K^N$. 
For a selected group of $n$ particles, the number of realizations of  configurations 
with $n$ particles in the selected $k$ cells and $N-n$ particles in the remainder 
$K-k$ cells is $k^n \cdot (K-k)^{N-n}$. For a totally random displacement of the particles
one obtains again the  Bernoulli distribution eq. (\ref{BERN1}).


Now we turn to the corresponding picture for bosonic systems. 
Bosons may be put into the same phase space cell in any number, without
limitation. Correspondingly $N > K$ is allowed and if so, there is a nonzero
probability for $n > k$, too. 

If one has $n$ indistinguishable particles, like bosons in quantum physics, then
only the occupation number of the cells labels a microstate.
The random placement by allowing multiple occupation in a cell counts for the
permutation of $n$ particles and $k-1$ separation marks. In this case the number
of configurations amounts to
\be
 W(n+k-1,n) \: = \: \binom{n+k-1}{n}.
\ee{QMULT}
In equilibrium this leads to the following P\'olya distribution:
\be
 Q_n \: = \: \frac{\binom{n+k-1}{n} \, \binom{(N-n)+(K-k)-1}{N-n}}{\binom{N+K-1}{N}}.
\ee{POLYAMULT}
In the dilute limit
\be
  W(n+k-1,n) \quad \xrightarrow[n \ll k]{} \quad \frac{k^n}{n!},
\ee{QMULTDILUTE}
and one obtains
\be
 Q_n \quad \xrightarrow[K \gg N]{n \ll k} \quad
 \binom{N}{n} \, \left(\frac{k}{K} \right)^n \, \left(1-\frac{k}{K} \right)^{N-n}.
\ee{MULTBERN}
The small subsystem limit on the other hand leads to the {\em negative binomial distribution}
\be
 Q_n \quad \xrightarrow[k \ll K]{n \ll N} \quad
 \binom{n+k-1}{n} \, p^n \, (1+p)^{-n-k},
\ee{MULTNBD}
with $p=N/K$. The expectation value of particles (connections) in the small subsystem 
amounts to $\exv{n}=kp=kN/K$ and the variance satisfies
\be
 \delta n^2 \: = \: \exv{n} \, \left( 1 + \frac{\exv{n}}{k}\right).
\ee{SMALLNBDVAR}

\section{Master Equation Approach}

One assumes that the elementary 
processes alter the state label $n$ only by one in a time step: either increasing or
decreasing it. The balance between these two processes is described then
by the stationary distribution, $Q_n$, which can be obtained based on
the knowledge of elementary growth and decay rates from the state with $n$
links (or particles in cell), $\mu_n$ and $\lambda_n$, respectively.
Due to the imposed conservation laws these rates are connected
by symmetry principles: e.g. $\mu_n(N-n)=\lambda_{N-n}(n)$. Often the
big environment  -- small subsystem limit, $n \ll N$ and $k \ll K$, 
in other cases the dilute limit $N \ll K$ and $n \ll k$ is considered.

In this case the evolution of $P_n(t)$ 
depends only on the state probabilities of having one more or one less quantum. 
The linearized version of this dynamics is described by
\be
 \dot{P}_n(t) = \lambda_{n+1} P_{n+1}(t) - \lambda_n P_n(t) + \mu_{n-1} P_{n-1}(t) - \mu_n P_n(t). 
\ee{MASTER} 
In this case the occurrence of state $n$
in a huge parallel ensemble (Gibbs ensemble) of systems is fed by both the $(n+1)\rightarrow n$ and
$(n-1)\rightarrow n$ processes and it is diminished by the reverse processes.
It is of special interest to investigate processes when $\lambda$ and $\mu$ are related
by symmetry principles, like time reversal invariance or 
by interchanging the role of subsystem and reservoir.

\subsection{Detailed balance distribution}

The general detailed balance solution of eq.(\ref{MASTER}), $Q_n$, we shall quote as the
{\em equilibrium distribution} (since this equation is homogeneous and linear in the $Q_n$-s
the overall normalization is not fixed by it). All $\dot{Q}_n$-s are zero only if
\be
  \lambda_{n+1} Q_{n+1} = \mu_n Q_n,
\ee{DET_BAL_PLUS}
from which it follows that also
\be
 \lambda_n Q_n = \mu_{n-1} Q_{n-1},
\ee{DET_BAL_MINUS}
annullating all evolution. Based on this observation, the detailed balance distribution
satisfies
\be
 Q_n = \frac{\prod_{i=0}^{n-1}\limits\mu_i}{\prod_{j=1}^{n}\limits\lambda_j } \, Q_0
\ee{DET_BAL}
and $Q_0$ can be obtained from the normalization condition $ \sum_n Q_n = 1$.
We note that besides the natural boundary conditions, 
\hbox{$P_{-1}=0$, $P_{\infty}=0$,} it follows the requirement $\lambda_0=0$ from the normalization
condition \hbox{$\sum_n\dot{P}_n=0$} upon eq.(\ref{MASTER}). 
This $\lambda_0$, however, does not appear in the stationary solution (\ref{DET_BAL}).

\subsection{Evolution of entropic distance: convergence}

Let us study the convergence to the equilibrium distribution, $Q_n$, starting from an arbitrary
$P_n(0)$ initial distribution. In order to follow the evolution of the ensemble, 
we use a general distance measure between two normalized distributions as follows:
\be
 \rho(P||Q)\:=\: \sum_n s\!\left(\frac{P_n}{Q_n} \right)\, Q_n\:=\: \sum_n s(\xi_n)\, Q_n.
\ee{distance1} 
Here $\xi_n(t)=P_n(t)/Q_n$ denotes the ratio to the detailed balance distribution.
Using the Jensen identity for a concave $s(\xi)$ (i.e. $s^{\prime\prime}(\xi) > 0$), 
\be
\sum_n s(\xi_n)\, Q_n \: \ge \: s\!\left( \sum_n \xi_n Q_n \right) \:=\:s(1),
\ee{Jensen}
one realizes that the natural choice $s(1)=\rho(Q||Q)=0$ ensures a proper
distance measure. This distance is not necessarily symmetric.
For $\rho(P||Q) \ne \rho(Q||P)$, it is often called just
a 'divergence'. We also demand that $s(\xi)\ne 0$ for any $\xi\ne 1$.
In this way any initial distance is shrinking until $\xi_n=1$ for all $n$.
The convergence is proven by $\dot{\rho}<0$ for $s\ne 0$.

This was demonstrated for a general, Monte Carlo type dynamics,
\be
 \dot{P}_n(t) \: = \: \sum_m w_{n\leftarrow m} P_m(t) \, - \, P_n(t) \sum_m w_{m\leftarrow n},
\ee{MONTECARLODYN}
by using a particular core function for the distance measure, $s(\xi)=\xi^2-1$,
in \cite{NARAYAN2001}.
It is, however, worth to be noted that any definition using a proper $s(\xi)$ core function
must lead to the same result. In the models discussed in the present paper
only those transition rates differ from zero for which $m= n\pm 1$.

Here the target distribution, $Q_n$, satisfies the {\em detailed balance} condition:
\be
 w_{n\leftarrow m} Q_m  \: = \: w_{m\leftarrow n} Q_n.
\ee{QDETBAL}
The time derivative of the distance defined above is then given by
\be
\dot{\rho}(t) \:=\: \sum_n s^{\prime}(\xi_n(t)) \: \dot{P}_n(t). 
\ee{rhorate1}
Using the general master equation (\ref{MONTECARLODYN}) one arrives at
\be
 \dot{\rho}(t) \: = \: \sum_{n,m}\limits s^{\prime}(\xi_n(t)) 
 \left( w_{n\leftarrow m} P_m(t)  \: - \: w_{m\leftarrow n} P_n(t)  \right).
\ee{RHODOT1}
From this point on the $t$-dependence of the quantities $P_n$, $\xi_n$ and $\rho$
are implicitely assumed.
Utilizing now the detailed balance condition, eq.(\ref{QDETBAL}), and replacing $P_n=\xi_nQ_n$
and $P_m=\xi_mQ_m$ one gains
\be
 \dot{\rho} \: = \: \sum_{n,m}\limits s^{\prime}(\xi_n)
 \, w_{m\leftarrow n} Q_n (\xi_m - \xi_n).
\ee{RHODOT2}
Finally in the above double sum the index notations $m$ and $n$ can be exchanged, and 
we arrive at
\be
 \dot{\rho} \: = \: \frac{1}{2} \, \sum_{n,m}\limits \, w_{m\leftarrow n} Q_n 
 \left(s^{\prime}(\xi_n) - s^{\prime}(\xi_m) \right) (\xi_m - \xi_n).
\ee{RHODOT3}
We apply now Lagrange's mid value theorem for equating the difference in
$s^{\prime}$ with a product of the interval length and the value of its derivative
in an internal point:
\be
 \dot{\rho} \: = \: - \frac{1}{2} \, \sum_{n,m}\limits \, w_{m\leftarrow n}Q_n \,
 s^{\prime\prime}(\xi^*_{nm}) \, \left(\xi_n-\xi_m\right)^2 \: \le \: 0.
\ee{RHODOT4}
with $\xi^*_{nm}=p\xi_n+(1-p)\xi_m$, $p\in[0,1]$.
Since $s^{\prime\prime}(\xi) > 0$, we arrive at the conclusion that
$\dot{\rho} < 0$ for any deviance from the targeted detailed balance distribution.
$\dot{\rho}=0$ can be achieved only if all $\xi_n=\xi_m=1$.

Finally we mention that another type of core function is also frequently used.
It is based on the Kullback-Leibler divergence, defined by $s(\xi)=-\ln \xi$.
As well for this as for the symmetrized definition,
$s(\xi)=(\xi-1)\ln\xi$, used in \cite{BIROSCHRAM}, there is always an approach
towards the detailed balance distribution.



\subsection{Symmetric decay and growth rates}

In the unified picture of moving stones between cells we label
as subsystem $k$ cells, and as environment the remaining $K-k$ cells. 
Having $n$ stones in the subsystem and the remaining $N-n$ in the environment,
$\lambda_n$ is the rate by which one of the $n$ stones is removed from the subsystem with $k$ cells, and 
$\mu_n$ is the corresponding rate for adding one stone. 
The simplest scenario is an independent removal of each stone, 
so the decay rate is proportional to the number of connections. 
For fermionic systems $n \le k$ and $N \le K$ conditions apply, while for bosonic
systems there are no such restrictions. 

The growth rate in equilibrium is given by
\ba
\lambda_n &=& \sigma \, n \, K,
\nl
\mu_n &=& \sigma  \, N  \, k.
\ea{SIMPLEST_NETWORK}
Here $\sigma$ parametrizes the overall speed of changes, being independent of $n$,
the factors $N$ and $K$ represent the symmetry condition with the environment.
Using eq.(\ref{DET_BAL}) these rates lead to a stationary state characterized by
the Poisson distribution.

A more sophisticated model considers a finite environment. In this way
the decay and growth rates reflect the size of the rest, $k$ for the growth
and $K-k$ for the the decay. Also the factor $N$ is replaced by
the reduced environment factor $N-n$. With such rates,
\ba
 \lambda_n &=& \sigma \, n \, (K-k),
\nl
 \mu_n &=& \sigma \, (N-n) \, k,
\ea{NET_RATES_BERNOULLI}
the stationary distribution is a Bernoulli one
\be
 Q_n \: = \: \binom{N}{n} \, v^n \, (1-v)^{N-n}
\ee{NET_STAT_BERNOULLI}
with $v=k/K$. In this case $\exv{n}=kN/K$ and indeed the $K\gg k$, $N \gg n$
limit leads back to the previous case discussed in eq.(\ref{SIMPLEST_NETWORK}).


Further decay and growth rates can be considered by  assuming 
random jumps of the stones between the cells, respecting the condition that 
in a cell maximum one stone can be put ({\em fermionic} case). At each time moment
a stone is randomly chosen and it is randomly repositioned in an empty cell.
The probability of choosing a given stone in the $k$ cells  is  $n/N$. 
The probability to choose now an empty cell that is not among the original $k$ cells is
given by the ratio $((K-N)-(k-n))/(K-N)$. Under such considerations one can write: 
\ba
 \lambda_n &=& w \, \frac{n}{N} \, \frac{(K-N)-(k-n)}{K-N},
\nl
 \mu_n &=& w \, \frac{N-n}{N} \frac{k-n}{K-N},
\ea{BIBI_RATES}
with $w=N(K-N)\sigma$.
Please note that these rates show a symmetry between the observed subsystem and 
its environment: one exchanges $\mu$ and $\lambda$ by changing 
$n \rightarrow (N-n)$ and $k \rightarrow (K-k)$. 

It is more transparent to express these rates
with the help of occupation and emptiness ratios as follows:
in accordance with the definition $p=N/K$ for the cell occupation ratio in the total system we use
$p_{{\rm sub}}=n/k$ and $p_{{\rm env}}=(N-n)/(K-k)$. Then $q_{{\rm sub}}=1-p_{{\rm sub}}$
and $q_{{\rm env}}=1-p_{{\rm env}}$ are the empty cell ratios  in the selected subsystem
and in the rest, respectively. Also $q=1-p$ the empty cell ratio in the total system. 
With these notations the growth and decay rates,
$\mu_n = \tilde{\sigma} q_{{\rm sub}} \cdot p_{{\rm env}}$ and 
$\lambda_n = \tilde{\sigma} p_{{\rm sub}} \cdot q_{{\rm env}}$
feature a symmetrically shaped formula with $\tilde{\sigma}=\sigma k(K-k)$. 
The stationary distribution, $Q_n$ depends only on their ratio 
\be
 \frac{\mu_n}{\lambda_n} \: = \: 
 \frac{p_{{\rm env}} \, (1-p_{{\rm sub}})}{p_{{\rm sub}} \, (1-p_{{\rm env}})}.
\ee{RATIOFORBERNOULLI}
One notes the occurence of the corresponding $(1-p)$ factors, analogous to the
Pauli-blocking factors in the Boltzmann--Uehling--Ulenbeck generalization of
the classical Boltzmann equation when dealing with fermions.

The stationary distribution according to eq.(\ref{DET_BAL}) in this case satisfies
\be
Q_n =   
\frac{ \left[ k(k-1)\ldots(k-n+1)\right] \cdot \left[N(N-1)\ldots(N-n+1)\right] }{\left[ 1\cdot 2 \cdot \ldots n \right] \cdot \left[ ((K-k)-(N-1))\ldots((K-k)-(N-n))\right] } 
 \, Q_0,
\ee{QnSATIS}
which, supplemented by 
\be
Q_0=\frac{(K-k)!(K-N)!}{(K-k-N)! K!} 
\ee{Q0v}
for achieving the correct normalization, leads finally
to the double binomial P\'olya formula (\ref{POLYA}).

Let us consider now indistinguishable bosonic stones. In such case the 
micro states of the system are given solely by the occupation numbers in the cells.
Two configurations are different if the occupation number of the cells are different. 
Let us assume that at each time-moment the occupation number of a randomly selected 
cell is lowered by one (if it is allowed, i.e. it is not 0), and the occupation number 
of another randomly selected cell is raised by one. In order to derive the $\lambda_n$ 
and $\mu_n$ rates, one has to take in consideration that only cells with 
occupation number different from $0$ can be selected. For a system with $n$ particles and $k$
cells the probability $q_0(n,k)$ that the occupation number in one selected cell is zero, 
can be computed as:
\be
 q_0(n,k) \: =
 \:  \frac{\binom{n+k-2}{n} }{\binom{n+k-1}{n}}=\frac{k-1}{n+k-1}
\ee{probab1}
The probability that the cell is occupied is then:
 \be
 p_0(n,k)=1-q_0(n,k)=\frac{n}{n+k-1}
\ee{probab2}
The average number of cells, $\overline{z}(n,k)$, that have nonzero occupation numbers is then
\be
 \overline{z}(n,k)=\sum_{j=0}^k j\cdot p_0(n,k)^j q_0(n,k)^{k-j} \, \binom{k}{j} \: = \: p_0(n,k)\cdot k.
\ee{average1}
With this in mind, the corresponding rates for the master equation become
 \ba
 \lambda_n &=& w \, \frac{\overline{z}(n,k)}{\overline{z}(N,K)} \, \frac{K-k}{K} \: = \: \alpha \, \frac{n}{n+k-1},
 \nl
 \mu_n &=& w \, \frac{\overline{z}(N-n,K-k)}{\overline{z}(N,K)} \, \frac{k}{K} \: = \: \alpha \, \frac{N-n}{(N-n)+(K-k)-1}
\ea{NBI_RATES}
with $\alpha=\tilde{\sigma} \, (K-N)(K+N-1)/K^2$.
\newline
The  stationary distribution corresponding to these rates writes as
\be
 Q_n = \frac{\left[N(N-1)\ldots(N-n+1)\right]\cdot\left[ k(k+1)\ldots (k+n-1)\right]}{\left[1\cdot 2\ldots n\right]\cdot\left[(K-k+N-1)\ldots (K-k+N-n)\right]} \, Q_0.
\ee{Q_NBIBI}
Using 
\be
 Q_0=\frac{(K-1)!}{(K-k-1)!} \, \frac{(N+K-k-1)!}{(N+K-1)!}
 \ee{origin1}
 we get exactly the P\'olya distribution from (\ref{POLYAMULT}).


For random networks it is fascinating to consider
{\em preferential attachment}. In this case the Matthias principle
is mostly applied: nodes with more connections have a better chance
to increase their connectivity. This translates to the construction of proper
transition rates as follows.
We apply a bosonic Boltzmann--Uhling--Uehlenbeck
type of modification of the rates in eq.(\ref{RATIOFORBERNOULLI}),
\ba
\lambda_n &=& \tilde{\sigma} \, p_{{\rm sub}} \,(1+p_{{\rm env}}) ,
\nl
\mu_n &=& \tilde{\sigma} \, p_{{\rm env}} \, (1+p_{{\rm sub}}).
\ea{BUU_RATES}
This definition incorporates the thresholded linear preference factor, $(k+n)$, to increase
the number of connections from $n$ to $n+1$ in $\mu_n$ and the symmetric
factor from the environment for $\lambda_n$. Using $k$ instead of $k-1$ this leads 
to the P\'olya distribution eq.(\ref{POLYAMULT}), while the infinite environment limit 
delivers the corresponding negative binomial distribution eq.(\ref{MULTNBD}).

To obtain a power-law tailed distribution \cite{BARABASI-PHYSICA,BIRO2006}  for $n$ 
using a balanced master equation, however, is way too artificial. 
Indeed, as we show in a forthcoming paper, power-law tailed 
stationary $n$-distributions occur much more naturally in far-equilibrium situations 
in open systems, when the growth dominates and the decay rate is set to zero.

\section{Application to particle spectra}

Particle multiplicity distributions, dominated by the bosonic
negative binomial one, are omnipresent in high energy accelerator experiments.
We briefly summarize here a simple argumentation, based on comparing phase space volumes.
Such distributions in the observed particle number $n$ may lead to certain,
power-law tailed shapes of the individual particle energy distributions.

Based on Einstein's argumentation the statistical factor in observing a hadron with energy $\omega$,
as emerging from a subsystem in a large quark-gluon soup with total energy $E$, 
is given by the phase space volume ratio \cite{BIRO-SIGMAPHI2014}: 
\be
 \rho(\omega) \: = \: \frac{\Omega(E-\omega)}{\Omega(E)}.
\ee{PHASVOLRATIO}
Considering ideal relativistic gases in a one-dimensional jet with randomly produced
$n+1$ particles, one simply assumes $\Omega(E)\sim E^n$. 
Therefore averaging over experimental events with fixed total energy, $E$, 
one predicts a statistical spectrum
\be
 \rho(\omega) \: = \: \sumi \left(1-\frac{\omega}{E}\right)^n \, Q_n.
\ee{PARTSPEC}
It is noteworthy that for the negative binomial distribution, eq.(\ref{MULTNBD}),
one obtains
\be
 \rho^{{\rm NBD}}(\omega) \: = \: \left(1+\frac{\exv{n}}{k} \, \frac{\omega}{E} \right)^{-k}.
\ee{SPECNBD}
This is a Tsallis--Pareto distribution \cite{TSALLIS-ALB} in the particle's energy, $\omega$,
with the temperature-like parameter $T=E/\exv{n}$, agreeing with the kinetic
definition of temperature, and $q=1+1/k$ Tsallis parameter. The latter quantity
measures the non-Poissonity \cite{IDEALGAS},
\be
 q = \frac{\exv{n(n-1)}}{\exv{n}^2}.
\ee{MEANINGOFTSALLISQ}
It is no wonder then that in the $q\to 1$ limit for dilute and small subsystems
the Poisson multiplicity distribution and a Boltzmann-factor, 
\be
 \rho^{{\rm POISSON}}(\omega) \: = \: e^{- \omega / T},
\ee{BOLTZMANNFACTOR}
emerge, parametrized only with the above kinetic temperature $T=E/\exv{n}$.

\section{Conclusion}

We have explored an analogy between (quantum) statistics of elementary particles
and node degree distribution in random networks. Both the stationary, near equilibrium
distributions and a simple master equation dynamics stabilizing those have been presented
in a common framework.

It has been shown that the simplest natural rates most commonly will lead to the
Poisson, Bernoulli and  P\'olya distributions. 
Networks resulting from an unrestricted random rewiring of the nodes,  should 
have their degree distribution in one of the above classes. 
In case multiple connections are allowed between two nodes the same distribution classes are expected.  
For a closed system with a random rewiring dynamics 
degree distributions different from the above ones hint for some sort of preferential 
mechanism in the rewiring dynamics.  
We have shown, however, that also for a special linear preferential 
rewiring dynamics one will get the P\'olya distribution. 

Our results for the network picture can be concluded as follows: 
(1) in case we limit the number of connections 
for a node the P\'olya type distribution is expected;  
(2) in the limit of large networks this distribution will tend to a
Bernoulli one; (3) in case the number of links per node is unrestricted one obtains
the negative binomial distribution.
However, in the limit of large networks  with a finite 
average connectivity per node,
we are driven to a Poissonian degree distribution.  
These distributions indeed occur 
in the degree statistics of random networks, and for large closed systems with a 
random rewiring dynamics usually the
Poisson degree distribution is the most common one \cite{NEWMAN}. 

For the particle production statistics the underlying $n$ distribution is indeed
very close to the negative binomial one. We have shown, that the statistical weight
for finding a given energy $\omega$ of a particle in this case follows the Tsallis--Pareto
distribution. This indeed has been confirmed by experimental particle spectra.
In the dilute limit the Poisson distribution emerges for $n$ and the spectra
reflect the thermal Boltzmann-Gibbs factor. The temperature in all the above cases
satisfies the familiar kinetic definition $T=E/\exv{n}$.

Although these results has been known for a while, the approach considered here can serve as
a concise classification scheme  unifying statistical approaches for 
random networks and high-energy particle production. The general result given in eq.(\ref{DET_BAL}) was
exploited here only for particular rates $\lambda_n$ and $\mu_n$. It is a challenging question
what stationary distributions may develop for nonlinear and/or unbalanced rates.

\section*{Acknowledgement}

This work has been supported by the Hungarian Scientific Research Fund OTKA,
supervised by the National Research, Development and Innovation Office NKFIH
(pro\-ject No.104260) and by a UBB STAR fellowship.
Discussions with A.~Telcs and Zs.~L\'az\'ar are gratefully acknoweldged. 
Z.~N\'eda acknowledges support from PN-II-ID-PCE-2011-3-0348 research grant.


\end{document}